# Scalability of VM Provisioning Systems

Mike Jones, Bill Arcand, Bill Bergeron, David Bestor, Chansup Byun, Lauren Milechin, Vijay Gadepally,
Matt Hubbell, Jeremy Kepner, Pete Michaleas, Julie Mullen, Andy Prout, Tony Rosa, Siddharth Samsi, Charles Yee,
Albert Reuther

Lincoln Laboratory Supercomputing Center
MIT Lincoln Laboratory
Lexington, MA, USA

*Abstract*—**Virtual machines and virtualized hardware have been around for over half a century. The commoditization of the x86 platform and its rapidly growing hardware capabilities have led to recent exponential growth in the use of virtualization both in the enterprise and high performance computing (HPC). The startup time of a virtualized environment is a key performance metric for high performance computing in which the runtime of any individual task is typically much shorter than the lifetime of a virtualized service in an enterprise context. In this paper, a methodology for accurately measuring the startup performance on an HPC system is described. The startup performance overhead of three of the most mature, widely deployed cloud management frameworks (OpenStack, OpenNebula, and Eucalyptus) is measured to determine their suitability for workloads typically seen in an HPC environment. A 10x performance difference is observed between the fastest (Eucalyptus) and the slowest (OpenNebula) framework. This time difference is primarily due to delays in waiting on networking in the cloud-init portion of the startup. The methodology and measurements presented should facilitate the optimization of startup across a variety of virtualization environments.**

I. INTRODUCTION

Despite the hype and ubiquity in recent years, the concept and technology of virtual machines have been around for over four decades. The first virtual machines were developed to share expensive mainframe computer systems among many users by providing each user with a fully independent image of the operating system. On the research front, MIT, Bell Labs, and General Electric developed the Multics system, a hardware and operating system co-design that featured (among many other things) virtual memory for each user and isolated program execution space [1]. Commercially, the pioneer in this technology was IBM with the release of System 360/67, which presented each user with a full System 360 environment [2]. Virtual machines went out of style in the 1980s as mainframes and minicomputers lost market share and personal computers became more widely accessible at work and at home.

The x86 architecture on which the PC revolution rode was not designed to support virtual machines, but in 1997, VMware developed a technique based on binary code substitution (binary translation) that enabled the execution of privileged (OS) instructions from virtual machines on x86 systems [16]. Another notable effort was the Xen project, which in 2003 used a jump table for choosing bare metal execution or virtual machine execution of privileged (OS) instructions [17]. Such projects prompted Intel and AMD to add the VT-x [19] and AMD-V [18] virtualization extensions to the x86 and x86-64 instruction sets in 2006, further pushing the performance and adoption of virtual machines.

Virtual machines have seen use in a variety of applications, but with the move to highly capable multicore CPUs, gigabit Ethernet network cards, and VM-aware x86/x86-64 operating systems, the use of virtual machines has grown vigorously, particularly in enterprise (business) computing. Historically, enterprise services were deployed to minimize complexity and conflict on each hardware server. The goal was to cost effectively isolate all of the services that were offered. Practically, this isolation means that one or a few enterprise service components were deployed on each hardware server. Thus, every time a new service was to be deployed or an existing service was to increase in scale, more hardware servers were added to the server closet, server farm, or data center. After years of expansion, this trend cannot be sustained by most organizations. Since most enterprise services are not particularly resource intensive (i.e., they do not consistently require more than a few percentages of CPU time, disk accesses, network calls, etc.), these enterprise services are very good candidates for deploying within virtual machines.

One of the important uses of virtual machines is the creation of test beds for new technologies. One example is the DARPA BRASS (Building Resource Adaptive Software Systems) program's evaluation platform. The goal of BRASS is to realize foundational advances in the design and implementation of survivable, long-lived complex software systems that are robust to changes in the logical or physical resources provided by their operational environment [3]. A key part of the BRASS program is its virtualized evaluation platform [20]. The platform is designed to assess the resiliency of new technologies. The platform architecture shown in Figure 1 relies heavily on a virtual machine infrastructure.

This material is based on work supported by DARPA under Air Force contract FA8721-05-C-0002. Any opinions, findings and conclusions or recommendations expressed in this material are those of the author(s) and do not necessarily reflect the views of DARPA.

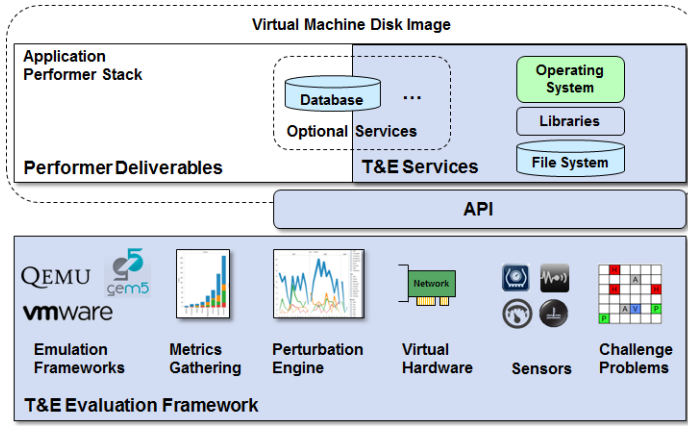

Fig. 1. Test platform for the DARPA BRASS program that relies heavily on a large-scale virtual machine framework running on a high performance computing cluster. The platform can run a variety of VMs from environments such as Qemu, VMware, and gem5. These VMs are launched so as to execute a range of challenge problems that are then perturbed and measured to test the resiliency of the technology under test.

VM performance is critical to the BRASS platform because resiliency experiments require that an application be tested under a wide range of conditions. One key aspect of VM performance is the time to start many VMs on a multicore system. This paper describes our measurements of the VM startup (launch request to VM userland entry) of three of the most widely used and mature Amazon Web Services' [4] compatible private cloud/infrastructure as a service (IaaS) platforms: OpenStack [11], OpenNebula [12], and Eucalyptus [13] (see Table I).

The selection of a VM infrastructure requires careful analysis of its features and performance. Several analyses have been done on the overhead of virtualization itself [15]. However, existing comparative VM research has been primarily focused on features. Few head-to-head benchmarks on controlled hardware have been done on VM orchestration and management. This paper describes an approach for accurately collecting startup time measurements.

The ability to quickly spin up and tear down virtual machines in an efficient manner is important in many situations of interest to the HPC community:

- Performing large-scale simulations of complex networks or the Internet.
- Isolating workloads with unconventional requirements – unusual/dated operating systems, libraries, applications, etc.
- Segregating user workloads to achieve heightened levels of security unavailable with discretionary access control systems.
- Rapidly provisioning prototype services to quickly demonstrate the feasibility or efficacy of a particular code while minimizing the cost of environment setup time.
- Setting up dynamically instanced, elastic database clouds to enable on-demand database computing. [5]

TABLE I: VM FRAMEWORKS TESTED

| | Framework | | |
|---|---|---|---|
| | *OpenStack* | *OpenNebula* | *Eucalyptus* |
| Hardware | HP ProLiant DL165 G7 32-core AMD Opteron 6274 96GB DDR3 RAM | | |
| Operating System* | CentOS 7.1 | CentOS 7.1 | CentOS 6.6 |
| Hypervisor | KVM | | |
| AWS-Compatible API | Yes | | |
| Startup Processes | 457 | 14 | 2 |
| Startup Threads | 834 | 102 | 1060 |
| VNC Console | Yes | Yes | Yes |
| Supports cloud-init | Yes | Yes | Yes |
| Image Format Used | QCOW2 | QCOW2 | RAW |

[a.] Latest officially supported CentOS Linux-based distribution.

A brief synopsis of the test platform configuration is presented in Table I, and the organization of the rest of the paper is as follows. Section II gives a high-level overview of the KVM [6] hypervisor, which all of the cloud management platforms tested relied upon. Section III explains how the tests were conducted and the manner in which the results were gathered. Section IV presents the results and a brief analysis of some of the reasons that could explain the disparity in the performance seen between these platforms. Section V summarizes this work and outlines potential future avenues of research based upon it.

## II. INSIDE THE KVM HYPERVISOR

The concept of operating systems was developed to present users with an abstraction to the system hardware, providing common services to the user, and managing and controlling shared resources that the hardware provides [7]. To host one or more operating systems within virtual machines on a host server, the virtual machine environment must provide the same capabilities.

Modern virtual machines are software abstractions of hardware architectures. Virtualization is accomplished by installing a hypervisor, which is also known as a virtual machine monitor (VMM). Much like the operating system provides abstracted services from the underlying hardware to the user, the hypervisor abstracts services from the underlying hardware. The hypervisor abstracts the underlying infrastructure, provides common services to one or more guest operating systems and their applications, and manages and controls the shared resources that the underlying infrastructure provides.

There are two primary modes in which a hypervisor can be implemented: type 1 (bare metal) and type 2 (hosted). Type 1 hypervisors typically execute instructions directly on the hardware, whereas type 2 hypervisors are installed as applications within another existing "host" operating system, and binary translation is used to map instructions from the guest operating system onto the host's hardware. This emulation layer, of course, comes with an associated performance cost that typically confines type 2 use to desktop, development, or rapid prototyping environments.

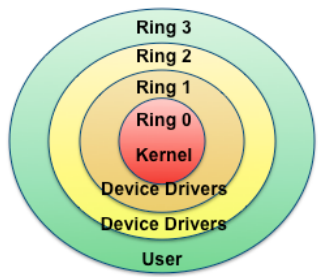

Fig. 2. Operating system privilege rings present in most modern hardware platforms and operating systems. x86 and x86-64 have four privilege rings as depicted, though Windows and Linux only use two (0 and 3) for supervisory and user privileges, respectively.

To manage and control the shared resources and provide common services to the guest operating systems, the hypervisor has its own kernel-like functionality as well as virtual device drivers that abstract the actual device drivers resident in the hardware. The kernel-like functionality manages and executes processes from the VMs. Much work in the research and commercial worlds has been done in this area, and CPU-intensive applications that are executed within VMs now pay a very small performance penalty for being executed within the VM. However, a greater performance penalty is paid by applications that perform much I/O through the hypervisor virtual device drivers such as network accesses and disk accesses. The more I/O that is required, the higher the performance penalty [8].

The other source of performance penalties for virtual machines is related to the execution of privileged instructions such as managing memory and I/O operations. Most hardware platforms (including x86 and x86-64) and most operating systems have multiple levels of privilege, often called privilege rings. The x86 and x86-64 architecture has four privilege rings numbered 0 through 3, as depicted in Figure 2. Most x86 and x86-64 operating systems, including Windows and Linux, only use rings 0 and 3 for supervisory and user privileges, respectively. In order for a ring 3 user application to execute privileged operations, such as requesting I/O or memory management services, the application must make a system call into the operating system kernel, where carefully vetted ring 0 supervisor-level kernel code executes the privileged request on behalf of the application.

With virtual machine environments, the hypervisor, virtual machine(s), and guest operating systems execute with ring 3 user privileges. A virtual machine OS does not have access to ring 0 supervisory privileges in the underlying hardware or host OS. In the x86 and x86-64 environment, this challenge was bridged by binary translation in VMware and jump tables in Xen. This overhead of privileged operations is the other main performance penalty in VMs.

The introduction of virtualization-enabling instruction in Intel VT-x and AMD's AMD-V in 2006 added ring -1 to the privilege ring, which is the supervisory ring for VMs. Ring -1 allowed hypervisors to use much less "glue" code to enable supervisory operations from virtual machines. Most hypervisors for x86 and x86-64 architectures now take advantage of these instructions, thus lessening the performance penalty for executing supervisory activities.

KVM is a type 1 or "bare metal" hypervisor, meaning that instructions from the virtual machine(s) are run directly on the x86 processor without a software emulation layer. The hypervisor has been entirely implemented as a pair of Linux kernel modules – one providing the core virtualization infrastructure and another for processor specific code. This architecture is depicted in Figure 3.

This approach allows for a very flexible implementation and carries many benefits. Many type-1 hypervisor implementations come bundled with a minimal operating system for the purpose of providing the tools and services necessary to run virtual machines and very little more. KVM has a complete, fully featured Linux operating system, with a standard kernel and available user libraries and binaries upon which additional tooling can be built.

Most of the operating system-level components needed for a hypervisor are already implemented in the standard Linux kernel – a process scheduler, a memory manager, networking and I/O subsystems – to allow a massive amount of code re-use. By implementing with new code only the VMM-specific features that are entirely orthogonal to existing kernel functionality as a separate kernel module and reusing existing structures where possible, KVM achieved inclusion into the mainline Linux kernel within a month of submission as a patch to the linux-kernel mailing list.

In addition to borrowing core systems code and structures from the Linux kernel, KVM relies upon a heavily modified version of QEMU [9], a mature open-source software machine emulator, to provide the user-land interface/process and virtual hardware (BIOS, including USB, SCSI, PCI, network drivers) upon which VMs can be launched.

The result is that the KVM hypervisor has increased performance and scalability, bereft of legacy code and architecture decisions predicated on previous-generation hypervisor engineering. Each virtual machine is simply another Linux process, and, in addition to being able to leverage all of the advanced features of current Linux scheduling, resource control, networking and security subsystems (control groups, mandatory access control frameworks, process/network namespaces, etc.), KVM is poised to make immediate use of any future advanced technologies developed as part of the standard Linux kernel engineering process [10].

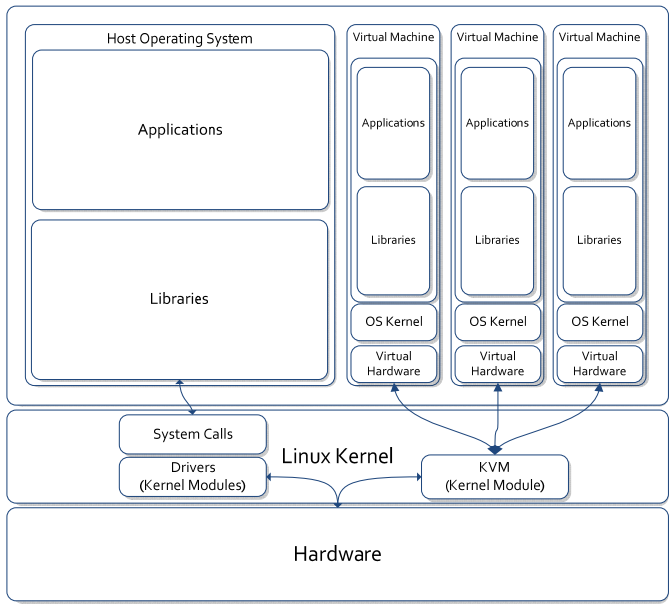

Fig. 3. Simplified KVM hypervisor architecture

### III. PERFORMANCE MEASUREMENT METHODOLOGY

Many VM infrastructures exist in large public cloud environments. Measuring the performance of VM infrastructure in this context could introduce added noise. To minimize uncertainty, performance measurements were performed on individual cluster nodes at the Lincoln Laboratory Supercomputing Center. These nodes could be fully isolated from extraneous processes. Four compute nodes were used: one to serve as a central test orchestrator, and three to host the cloud frameworks under test. Each compute node was an HP ProLiant DL165 G7 with dual socket AMD Opteron 6274 processors, 96GB of RAM per node, and four 1TB 2.5" hard drives.

A test orchestration server running Fedora Core 20 was set up to serve as a central point from which to launch test runs and collect the resulting data. The Apache web server was used to collect timing data from callback HTTP requests sent from each of the launched VMs once provisioning was complete.

The three VM frameworks under test were installed and configured with their respective default parameters; in each case, the latest officially supported version of CentOS was used as the base operating system; for OpenStack [11] and OpenNebula [12], there was a set of official CentOS 7.1 packages, whereas for Eucalyptus [13], the latest supported distribution at the time of testing was CentOS 6.6.

KVM was selected as the underlying hypervisor for all tests because it was uniquely supported by all three platforms chosen.

Each full sequence of VM launches was preceded with an HTTP request from the server running the cloud management software (Eucalyptus, OpenStack or OpenNebula) to the test orchestration server to indicate the beginning of the test run.

The VM launched was a slightly modified minimal install of Fedora Cloud 22 using their official cloud image. This VM was modified to accurately report when the server has been successfully provisioned and booted. More specifically, as the final step in the cloud-init provisioning of the virtual machine (just prior to handoff to the remainder of the systemd boot process), a final HTTP request is made to the test orchestration server. This final request using the Linux command line tool 'curl' with a specially crafted URL assists in identifying the particular instance in question, and indicates that the virtual server has been successfully provisioned and booted. The workflow for this process is described in Figure 4.

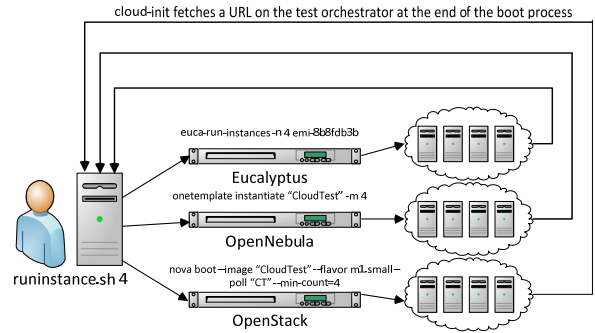

Fig. 4. Orchestration workflow depicts the launch of a test run, its deployment across a series of servers running cloud management frameworks and the return calls from the launched virtual instances back to a web server running on the orchestration server from which the test run was launched.

The measurements consisted of taking the difference between the start time and the final time as reported in the web server log. The start time was taken from the initial HTTP request at the launch of the test. The final time was taken from and the final HTTP request of the VM that was last provisioned during that test.

Each of the three platforms had a unique method of requesting a numbered batch of VMs be started from a particular "template." The commands used to launch the VM sets for each of the systems being evaluated are provided in Figure 5.

```
euca-run-instances -n vm-count vm-image-name

nova boot --image vm-image-name --flavor m1.small --
poll vm-name --nic network-uuid --min-count vm-count

onetemplate instantiate "vm-image-name" --name "vm-
name" -m vm-count
```

Fig. 5. Linux commands, launched from a script on the VM framework server itself, used to provision a numbered batch of virtual machines. The first line is for Eucalyptus, the second for OpenStack, and the third for OpenNebula.

### IV. STARTUP PERFORMANCE RESULTS

Since all three cloud orchestration platforms were built upon the same underlying technologies (QEMU paired with KVM as the hypervisor, running on CentOS Linux) run on identical hardware, launching an identical, stripped-down image of Fedora 22 Cloud, we were able to measure the ability of these three software stacks to rapidly spin up and tear down instances of virtual machines.

Figure 6 shows the provisioning test results. Each line represents a different VM provisioning system. While the results show a significant disparity between launch times across the spectrum of tests performed, a great deal of the delta in startup time between the three platforms can be attributed to delays in their handling of the cloud-init provisioning tool upon which the test relied.

OpenNebula suffered the most heavily from cloud-init delays, taking upwards of 120 seconds to start a single instance. Most of this delay was spent waiting on networking in cloud-init. As cloud-init is the leading industry standard for bootstrapping virtual machines in the public and private clouds, with the support of both enterprise cloud platforms such as Amazon Web Services and enterprise Linux distribution vendors like RedHat and Ubuntu (among many others), we feel that it serves as a fair point at which to measure "completion" of VM setup as it pertains to a VM provisioning system.

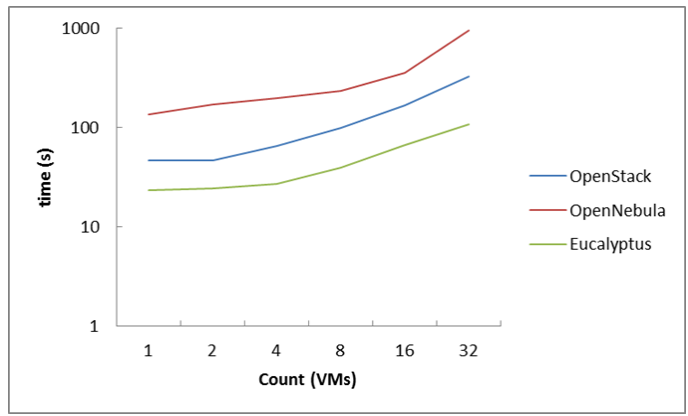

Fig. 6. Total time taken to provision the entire set of VMs requested.

## V. SUMMARY AND FUTURE WORK

With this benchmark of open-source cloud VM provisioning systems, we have demonstrated that even when the technologies underpinning the current management platform offerings are close to identical from hardware to user-land, there is a large degree of inconsistency in VM launch performance, as perceived from a user standpoint, depending upon which platform is chosen.

Using virtualization technology in HPC is a compelling option for certain workloads with unique or unusual requirements that preclude configuring an HPC cluster to accommodate them [14].

A 10x performance difference is observed between the fastest (Eucalyptus) and the slowest (OpenNebula) framework. This time difference is primarily due to delays in waiting on networking in the cloud-init portion of the startup. The methodology and measurements presented should facilitate the optimization of startup across a variety of virtualization environments.

Future work will include further scaling of the VM platforms examined in this paper by launching each platform across multiple compute nodes with a more complex networking configuration, testing various contention ratios of physical cores versus VM instances launched, and including additional, comparable cloud platforms, such as Apache CloudStack, which were not evaluated in this experiment.

ACKNOWLEDGMENTS

The authors wish to acknowledge the following individuals for their contributions: Jeff Hughes, Suresh Jagannathan, and David Martinez.